\documentclass[a4paper]{jpconf}
\usepackage{graphicx}
\usepackage{amsmath}
\begin{document}
\title{Chirality-spin separation in the Hubbard model on the kagome lattice}

\author{Masafumi Udagawa and Yukitoshi Motome}

\address{Department of Applied Physics, University of Tokyo,
7-3-1 Hongo, Bunkyo-ku, Tokyo, Japan}

\ead{udagawa@ap.t.u-tokyo.ac.jp}

\begin{abstract}
Effect of geometrical frustration in strongly-correlated metallic region 
is studied for the Hubbard model on the kagome lattice at half filling
by a cluster extension of the dynamical mean-field theory 
combined with a continuous-time auxiliary-field quantum Monte Carlo method. 
We find that the electron correlation enhances the spin chirality in
both vector and scalar channels. 
The chirality grows as decreasing temperature 
and exhibits a peak at a low temperature, 
indicating a new energy scale under strong correlation. 
The peak temperature is considerably lower than 
that for the local spin moment, namely,
the characteristic temperatures for the chirality and the local moment 
are well separated. 
This is a signature of separation between spin and chiral degrees of freedom 
in the correlated metallic regime under geometrical frustration.
\end{abstract}

\section{Introduction}
For a long time, considerable attention has been focused on the role of geometrical frustration in 
localized spin systems. It has been discussed that 
large degeneracy due to geometrical frustration opens a way to realize  
non-trivial ground states, such as spin-liquid state \cite{rf:Anderson1, rf:Anderson2},
by suppressing conventional long-range ordering. 
It has also been clarified that geometrical frustration can induce an exotic long-range order 
composed of some higher rank objects, 
such as spin chirality \cite{rf:Miyashita} and spin quadrupole \cite{rf:Tsunetsugu}.

On the other hand, much less is known about the effect of geometrical frustration 
in itinerant electron systems. 
Recently, several interesting phenomena have been found in itinerant systems 
and an importance of the geometrical frustration has been pointed out.
A typical example is heavy-fermion behavior found in transition metal compounds 
such as LiV$_2$O$_4$ \cite{rf:Kondo}, Y(Sc)Mn$_2$ \cite{rf:Wada} and 
$\beta$-Mn \cite{rf:Shinkoda}. In contrast to the heavy-fermion rare-earth compounds, 
these materials have no explicit localized moments, and therefore, 
it is difficult to explain their heavy-fermion behavior by the conventional 
Kondo mechanism based on the interaction between conduction electrons and localized spins.
A candidate for the origin of the heavy-fermion behavior is
a common feature among these transition metal compounds ---
geometrical frustration in the underlying lattice structure.
In fact, in the related insulating materials, it has been reported that magnetic transition temperatures are strongly
suppressed compared to the Curie-Weiss temperatures because of the frustration \cite{rf:Ueda}.
Furthermore, characteristic magnetic fluctuations under a suppression of long-range ordering 
are observed in these materials,
and their relation to the heavy-fermion behavior has attracted much interests
\cite{rf:Lee,rf:Ballou,rf:Nakamura}. 

In general, in a weakly-correlated metal, the electronic state is well described by a Slater determinant of 
single-particle states which extend over the entire system. In this case, 
physical properties of the system will be rather insensitive
to the local lattice structure, i.e., whether or not the lattice is geometrically frustrated.
The issue is how this picture is modified when the system enters into a strongly-correlated regime
where the electrons tend to be localized due to strong electron interaction. 
In particular, in the vicinity of the Mott transition, spin moments grow with antiferromagnetic correlations 
between neighboring sites, and hence, it is expected that these moments suffer from the frustration 
and lead to some exotic behavior as seen in the localized spin systems. 
It is intriguing to examine how the geometrical frustration affects 
the electronic state through the interplay between spin and charge degrees of freedom. 

Motivated by these considerations, in this paper, we investigate the effect of geometrical frustration in correlated metallic region 
for a simple model, the Hubbard model on the kagome lattice. 
In particular, we will focus on the behavior of spin chirality in comparison with that of the original spin degree of freedom. 

\section{Model and Methods}
As a minimal model including both geometrical frustration and electron correlation,
we consider the Hubbard model on the kagome lattice, whose Hamiltonian is given by
\begin{eqnarray}
\mathcal{H} = -t \sum\limits_{\langle i,j\rangle,
 \sigma} \bigl(c^{\dagger}_{i\sigma}c_{j\sigma} +  {\text{h.c.}}\bigr) +
 U\sum\limits_i n_{i\uparrow}n_{i\downarrow} - \mu\sum\limits_{i\sigma}n_{i\sigma},
\end{eqnarray}
in the standard notations.
We consider only the nearest-neighbor site hopping 
on the kagome lattice shown in the left panel of Fig.~\ref{lattice_to_imp}, 
and set $t=1$ as an energy unit hereafter.
The chemical potential $\mu$ is controlled so that the system is at half filling
(one electron per site on average). 
The Mott transition and magnetic fluctuations near the transition have been recently studied for this model by several theoretical methods \cite{rf:Imai, rf:Bulut, rf:Ohashi}.

To investigate the correlated metallic region of the model (1), we use a cluster extension of the 
dynamical mean-field theory \cite{rf:Kotliar}, which enables to describe the Mott transition 
with including spatial fluctuations within a finite-size cluster. 
In this method, the original lattice problem is
mapped to a cluster impurity problem: 
In the present study, in order to take account of the effect of geometrical frustration,
we consider a 3-site cluster (Fig.~\ref{lattice_to_imp}). We extended the cluster up to 12 sites and confirmed that 
the spin correlation hardly develops beyond the nearest-neighbors 
in the parameter range of the following calculations (not shown). 
To solve the impurity problem, we employ the continuous-time auxiliary-field quantum Monte Carlo method
\cite{rf:Gull}, which requires much smaller computational cost compared
with the conventional Hirsch-Fye algorithm.

In the next section, we will show the results for the density of states, 
the spin chirality and the local spin moment. 
The density of states (DOS) $\rho(\omega)$ is calculated from 
the one-particle local Green's function, $G_{j\sigma}(\tau)=-\langle T_{\tau}c_{j\sigma}(\tau)c^{\dag}_{j\sigma}(0)\rangle$,
($j=1-3$ corresponds to the cluster sites in Fig.~\ref{lattice_to_imp}) 
by using $\rho(\omega)=-\frac{1}{\pi}\text{Im}G_{j\sigma}(\omega+i\delta)$.
We use the maximum entropy method
for the analytic continuation. 
Note that $\rho(\omega)$ does not depend on either $j$ or $\sigma$
within the paramagnetic solution assumed here. 
For the calculations of magnetic properties, 
we define the local spin at cluster site $j$ as 
${\mathbf s}_j^\nu\equiv\frac{1}{2}\sum_{s,s'}c^{\dag}_{js}(\mbox{\boldmath $\sigma$}^\nu)_{ss'}c_{js'}$, with the $\nu$-th component of Pauli matrix, $\mbox{\boldmath $\sigma$}^\nu$ ($\nu=x,y,z$). 
By using ${\mathbf s}_j$, we calculate the vector spin chirality, ${\mathbf K}_v\equiv\frac{2}{3\sqrt{3}}
({\mathbf s}_1\times{\mathbf s}_2 + {\mathbf s}_2\times{\mathbf s}_3 + {\mathbf s}_3\times{\mathbf s}_1)
$, and the scalar spin chirality, ${\mathbf K}_s\equiv({\mathbf s}_1\times{\mathbf s}_2)\cdot{\mathbf s}_3$.
In the following, we present the squared moments of the vector chirality $K_{v}^2\equiv\langle{\mathbf K}_v^2\rangle$ and of the scalar chirality $K_s^2\equiv\langle{\mathbf K}_s^2\rangle$ 
as well as the local spin moment $s^2\equiv\langle{\mathbf s}_j^2\rangle$
($s^2$ is independent of $j$). Here, the bracket represents the
statistical average taken in the grand-canonical ensemble at temperature $T$. 
$s^2$, $K_v^2$ and $K_s^2$ are two-, four- and six-body equal-time correlation functions, 
respectively, and obtained by applying the Wick's theorem for each Monte Carlo sampling.

\section{Results}
\begin{figure}[h]
\begin{minipage}[t]{17pc}
\includegraphics[width=15.4pc]{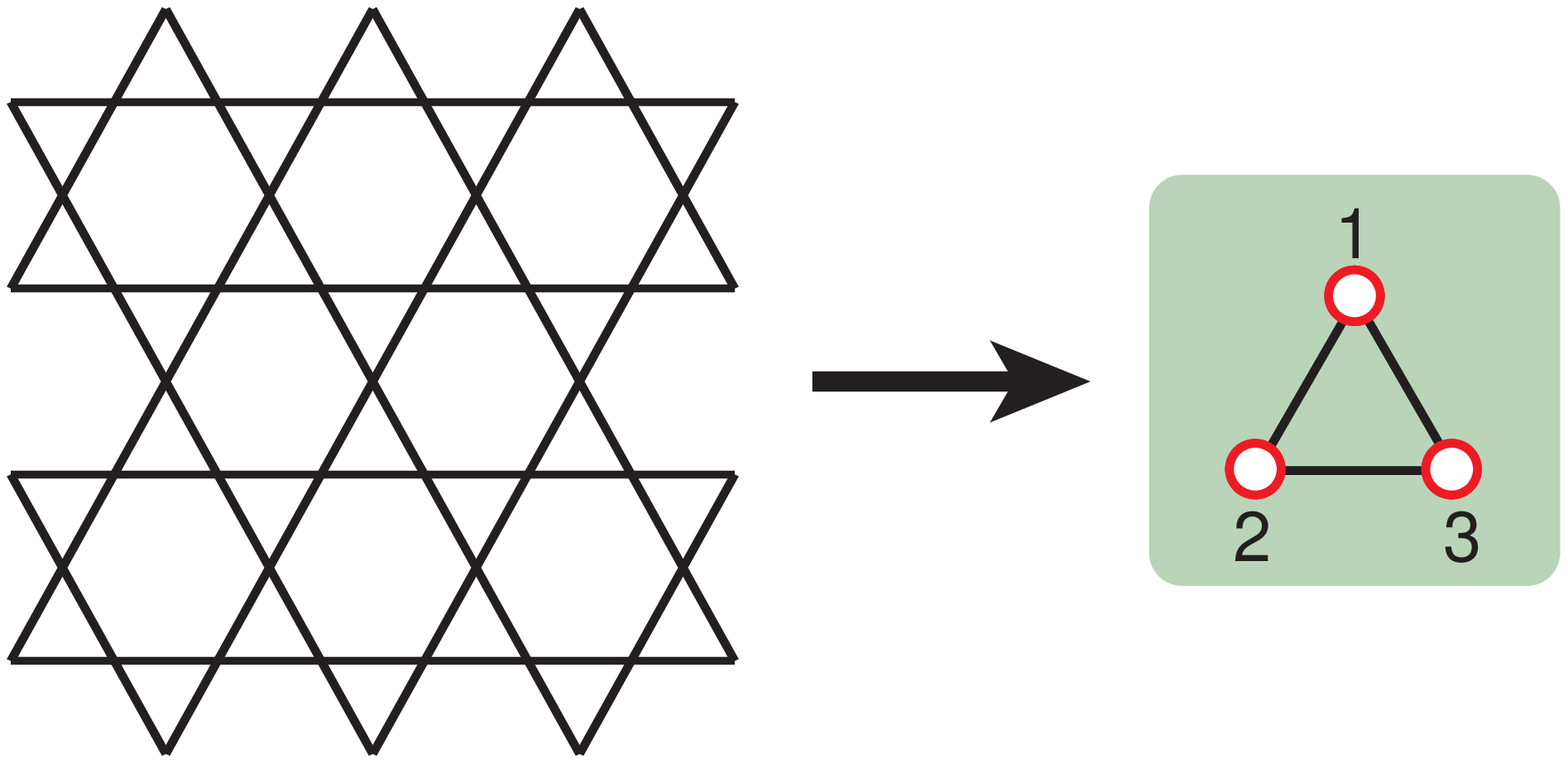}
\caption{\label{lattice_to_imp}Schematic picture of the mapping of the Hubbard model on the kagome lattice to an effective cluster-impurity Anderson model.}
\end{minipage}\hspace{2pc}%
\begin{minipage}[t]{16pc}
\includegraphics[width=15pc]{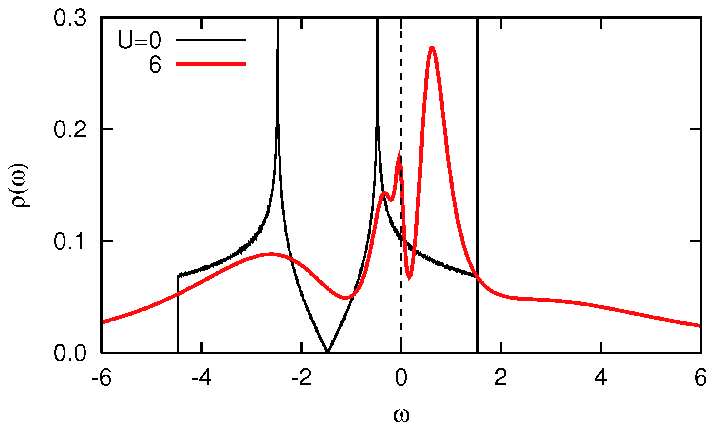}
\caption{\label{DOS} DOS at $U=6$ and $T=0.05$ (red) in comparison with the result at $U=0$ (black). We set the chemical potential to be $\omega=0$.}
\end{minipage} 
\end{figure}

First, we present the result of DOS at $U=6$ and $T=0.05$ in Fig.~\ref{DOS}. For comparison, we also show the non-interacting DOS at $U=0$. 
At $U=0$, $\rho(\omega)$ structures, two van-Hove singularities ($\omega \simeq -0.47$ and $-2.47$) and a flat band ($\omega \simeq 1.53$). 
At $U=6$, the system is still in the metallic region with finite DOS at the Fermi level $\omega=0$, 
whereas $\rho(\omega)$ shows several characteristic features of strong correlation effect. 
One is the strong renormalization of the energy scales; the three peaks around $\omega=0$ are considered as renormalized structures of the van-Hove singularities and the flat band. 
Another feature is the formation of the Hubbard bands which are observed as two broad humps at $\omega\sim\pm3$. 
We note that the result is consistent with the previous one obtained 
by a similar method but with using the conventional Hirsch-Fye algorithm \cite{rf:Ohashi}. 
The previous study revealed that the critical value of $U$ for the Mott transition is about 8.2. 
Below we show the results for the spin chirality degrees of freedom 
in this correlated metallic region at $U=6$.

\begin{figure}[h]
\begin{minipage}[t]{16pc}
\includegraphics[width=15.3pc]{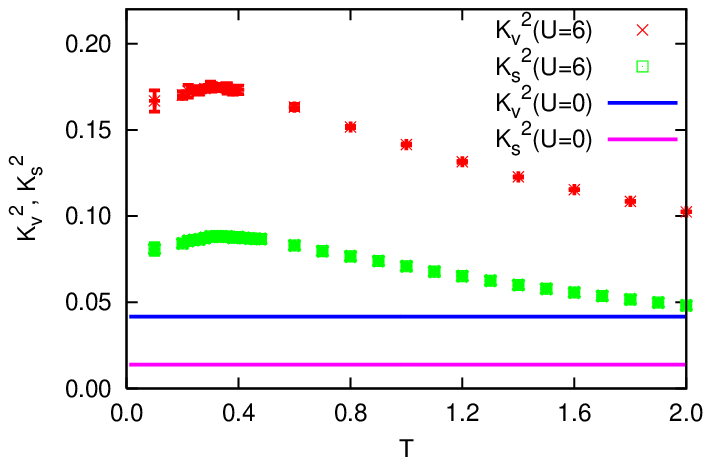}
\caption{\label{KatU0and6}Temperature dependence of $K_v^2$ and $K_s^2$ at $U=6$ and $U=0$.}
\end{minipage}\hspace{2pc}
\begin{minipage}[t]{16pc}
\includegraphics[width=17pc]{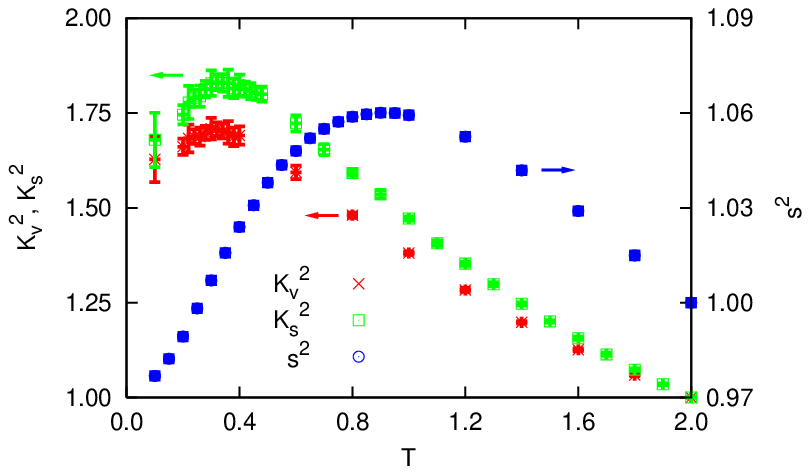}
\caption{\label{SandK}Temperature dependence of $K_{v}^2$, $K_s^2$ and $s^2$ at $U=6$. All the data are normalized by the value at $T=2.0$.}
\end{minipage} 
\end{figure}

Figure \ref{KatU0and6} shows the temperature dependence of the spin chiralities $K_{v}^2$ and $K_s^2$. 
At $U=0$, both $K_v^2$ and $K_s^2$ are small and featureless, in spite of the peculiar $\omega$ dependence of the DOS. 
The deviation from the high temperature limits [$K_v^2(T\rightarrow\infty)=1/24$ and $K_s^2(T\rightarrow\infty)=3/256$] is within 0.03 \%
for the entire temperature range. In contrast, 
at $U=6$, both $K_v^2$ and $K_s^2$ are largely enhanced compared with the non-interacting case. Furthermore, as decreasing $T$, they grow gradually and exhibit a peak at a low temperature 
$T_{K}^* \simeq 0.3$. This temperature $T_K^*$ characterizes a 
new energy scale related with the spin chirality degrees of freedom.  

In Fig.~\ref{SandK}, we compare this behavior of the chiralities with the local spin moment $s^2$. $s^2$ also shows a broad peak, 
but the peak temperature $T_s^*\simeq0.9$ is considerably higher than $T_K^*$.
In the intermediate temperature $T_K^* < T < T_s^*$, 
the chiralities are enhanced while the spin moment is suppressed as decreasing $T$. 
This contrastive behavior is surprising because of the following reasons. First of all, by definition, 
the spin chiralities are composed of the products of spin operators. 
In fact, their behavior is tightly related with that of the spin moment 
in most of the localized spin systems. Furthermore, in the present itinerant system, 
the decrease of $s^2$ is anticipated to reduce the spin chirality moments
since it corresponds to the increase of 
doubly-occupied or empty sites: 
The spin chiralities have largest values when they operate on the
3-site triangle which includes no doubly-occupied or empty site.
Therefore, the apparent separation between the spin and chirality found in Fig.~\ref{SandK} 
is highly non-trivial. 

\section{Summary and Concluding Remarks}
We have studied the spin and chirality moments 
for the Hubbard model on the kagome lattice at half filling 
by applying the cluster dynamical mean-field theory
combined with the continuous-time quantum Monte Carlo method. In metallic region under strong electron correlation, we obtained the following results. 
(1) The spin chirality is strongly enhanced in both vector and scalar channels,
compared with the non-interacting case. 
It grows as decreasing $T$ and exhibits a peak at a low temperature $T_K^*$, implying 
a new energy scale due to the electron correlation. 
(2) The local spin moment also shows a peak, 
but the peak temperature $T_s^*$ is substantially higher than $T_K^*$.
In the temperature range between $T_K^*$ and $T_s^*$, 
the spin and chirality show opposite temperature dependences. 
These results suggest a separation of the spin and chirality degrees of freedom
in the correlated metallic region in this strongly frustrated system.

The origin of the chirality-spin separation deserves further study.
It is also interesting to clarify how the electronic state is affected
by the appearance of the new energy scale $T_K^*$ related with
the chirality. Another issue is a possibility of some instabilities in the chiral sector
by allowing symmetry breaking within the cluster dynamical mean-field framework. 
Systematic study in a wide region of $U$ and $T$ is now in progress. 
 
 We would like to thank S. Sakai for fruitful discussions. We also  acknowledge S. Onoda and H. Tsunetsugu for helpful comments.
This work was supported by Grant-in-Aid for Scientific Research on Priority Areas 
(Nos. 17071003, 19052008), Grant-in-Aid for Young Scientists(B) (No. 21740242), Global COE Program ``the Physical Sciences Frontier" and
by the Next Generation Super Computing Project, Nanoscience Program, MEXT, Japan.

\end{document}